\documentclass [12pt]{article}
\usepackage{amssymb,amsmath}
\usepackage{amsthm}
\usepackage{graphicx}

\usepackage[cp1250]{inputenc}
\usepackage[active]{srcltx}

\long\def\comment#1{{}}

\newtheorem{thm}{\bf Theorem}

\newtheorem{rem}{\bf Remark}

\theoremstyle{definition}
\newtheorem{ex}{\bf Example}

\title{\textbf{On the Convergence of the Pairwise Comparisons Inconsistency Reduction Process}}  
\author{W.W.~Koczkodaj \footnote{Computer Science, Laurentian University,
935 Ramsey Lake Road, Sudbury,
ON P3E 2C6 Canada, wkoczkodaj@cs.laurentian.ca}
\and J.~Szybowski \footnote{AGH University of Science and Technology,
Faculty of Applied Mathematics,
al. Mickiewicza 30, 30-059 Krakow, Poland, szybowsk@agh.edu.pl}}

\begin{document}
\maketitle             
\begin{abstract}
This study investigates a powerful model, targeted to subjective assessments, based on pairwise comparisons. It provides a proof that a distance-based inconsistency reduction transforms an inconsistent pairwise comparisons (PC) matrix into a consistent PC  matrix which is generated by the geometric means of rows of a given inconsistent PC matrix.  The distance-based inconsistency indicator was defined in 1993 for pairwise comparisons. Its convergence was analyzed in 1996 (regretfully, with an incomplete proof; finally completed in 2010). However, there was no clear interpretation of the convergence limit which is of considerable importance for applications and this study does so. This study finally ends the ongoing (since 1984) dispute on the approximation method for the inconsistent pairwise comparisons. The convergence limit is the vector of geometric means. It is not the principal right eigenvector of a given PC matrix.
\end{abstract}


\section{Basic concepts of pairwise comparisons}
\label{sec:bc}

In modern science, we compare entities in pairs even without realizing it. For example, when someone asserts: ``I am 18 years old,'' one year is compared to the length of his/her life. It is a pair: one entity (year) is a unit and another entity is magnitude the life duration. When we have no unit (e.g., for software reliability), we may consider construction of a pairwise matrix to express our assessments based on relative comparisons of its attributes (such as safety or reliability). 

Pairwise comparisons is primarily about the information fusion from the most primitive pairwise (hence binary) to a single measure of the goal (such as importance or public safety).

In this study, we assume that {\em pairwise comparisons} {\em matrix} (PC matrix here) is a square matrix $M=[m_{ij}]$, $n \times n$, such that $m_{ij}>0$ for every $i,j=1, \ldots ,n$. PC matrix $M$ is called {\em 
reciprocal} if $m_{ij} = \frac{1}{m_{ji}}$ for every $i,j=1, \ldots ,n$
(in such case, $m_{ii}=1$ for every $i=1, \ldots ,n $). However, the blind wine testing may result not only in the lack of reciprocity but even in the lack of 1's on the main diagonal since, comparing the same wine to itself (especially at the end of the tasting day), may not necessary be correct. Projects may be compared in different locations by different assessors due to the Internet. In such situation, it is even anticipated that some (if not most) assessments may not be reciprocal since one team of experts may assess project $A$ to be more important for development than $B$ and another team may assign the higher priority to $B$ without even realizing that $A/B$ was done so they could simply use the value $1/(A/B) $.
\\

Let us assume that:

\begin{displaymath}
M = \begin{bmatrix}
1 & m_{12} & \cdots & m_{1n} \\ 
\frac{1}{m_{12}} & 1 & \cdots & m_{2n} \\ 
\vdots & \vdots & \vdots & \vdots \\ 
\frac{1}{m_{1n}} & \frac{1}{m_{2n}} & \cdots & 1
\end{bmatrix},
\end{displaymath}

\noindent where $m_{ij}$ expresses a relative preference of an entity $E_i$ over $E_j$. An entity could be any object, attribute of it or a stimulus.

A pairwise comparisons matrix $M$ is called {\em consistent} (or {\em transitive}) if 
$$m_{ij} \cdot m_{jk}=m_{ik}$$ for every $i,j,k=1,2, \ldots ,n$. \\

We will refer to it as a ``consistency condition.'' In layman's terms, when we have three entities: $A, B, C$, then $A/B \cdot B/C$ must yield the same result as the comparison $A/C$ but often, it does not when these three comparisons are carried independently. When the comparisons or entities (safety, reliability, etc.) are subjective, the inconsistency is unavoidable in practice. In fact, the lack of inconsistency in such case may be suspicious but it does not mean that the inconsistency is desirable or that it should be tolerated. The ``GIGO rule'' (GIGO stands for ``garbage-in, garbage-out'')  in the field of computer science and information technology refers to the fact that the output quality depends on the quality of input data. It is not only common sense but the famous ``modus ponens'' (Latin for ``the way that affirms by affirming'') in propositional logic which indicates the seriousness of the problem. 
It can be viewed as ``P implies Q; P is asserted to be true, so therefore Q must be true'' but it is the ``Generalized Modus Ponens'' in fuzzy logic which has a great application here. It generalizes the classical inferential rule of modus ponens and can be summarized (in layman's terms) as:

\begin{quote}
If the premise $A'$ is slightly different from $A$, \\
then the conclusion $B'$ is slightly different from $B.$
\end{quote}

While every consistent matrix is reciprocal, the converse is false in general. If the consistency condition does not hold, the matrix is inconsistent (or intransitive). In at least four studies \cite{KSB1939, H1953, GS1958, S1961},  conducted between 1939 and 1961, the inconsistency in pairwise comparisons was defined and examined.
   
Consistent matrices correspond to the ideal situation in which there are the exact values $E_1, \ldots , E_n$ for the stimuli. 
Then the quotients $m_{ij}=E_i/E_j$ form a consistent matrix. 
The vector $s=[E_1, \ldots E_n]$ is unique up to a multiplicative constant. 
The challenge of the pairwise comparisons method comes from the lack of consistency of the pairwise comparisons matrices which arises in practice (while, as a rule, all the pairwise comparisons matrices are reciprocal). Given a $n \times n$ matrix $M$, which is not consistent, the theory attempts to provide a consistent $n \times n$ matrix $M'$ which differs from matrix $M$ ``as little as possible''. 

The matrix: $M= E_i/E_j$ is consistent for all (even random) values $v_i$. It is an important observation since the consequence of it is that the approximation is really a problem of the distance minimization. For the Euclidean norm, the vector of geometric means (which is equal to the principal eigenvector for a consistent PC matrix) is the one which generates it. 
Saaty's seminal study \cite{Saaty77} had a considerable impact on the pairwise comparisons research. 
It has strongly endorsed the use of eigenvector, corresponding to the principal eigenvalue, for approximation of a given inconsistent but reciprocal PC matrix. 

In \cite{Saaty77}, a {\em consistency index} was proposed as a scaled deviation of the difference between matrix size and the principal eigenvalue:

$$CI= {\lambda_{max}-n \over {n-1}}$$

\begin{flushleft}
Conclusions of \cite{XDXW2008} include:
\end{flushleft}
\begin{quotation}
``in this paper, by simulation analysis, we obtain the following
result: as the matrix size increases, the percent of the matrices
with acceptable consistency ($CR \leq 0.1$), decrease dramatically,
but, on the other hand, there will be more and more contradictory
judgments in these sufficiently consistent matrices. This paradox
shows that it is impossible to find some proper critical values
of CR for different matrix sizes. Thus we argue that Saaty's
consistency test could be unreasonable.''
\end{quotation}

Authors of the above text use CR (for ``Consistency Ratio'') while CI (for ``Consistency Index'') is more commonly used.


In \cite{KS2014a}, two counter-examples, with the mathematical reasoning and proofs, demonstrated that the eigenvalue-based inconsistency tolerates an error of any arbitrarily large value (e.g., 1,000,000\% or whatever our imagination calls for). The eigenvalue-based inconsistency fails rather simple axioms proposed in \cite{KS2014a}. 
The same reasoning is also applicable to any other panoptic inconsistency indicator. By panoptic, we understand any type of a global characteristic including ``averaging'' statistics, outcomes from regression-based models or expressions using eigenvalues. 
The distance-based inconsistency (introduced in \cite{Kocz93}) is consistent with the simple axiomatization proposed in \cite{KS2014a}. 

\section{Inconsistency in pairwise comparisons }
\label{sec:ic}

As pointed out earlier, given an inconsistent PC matrix, a number of theories have attempted to approximate it with a consistent matrix that differs from matrix $M$ ``as little as possible.'' 
Needless to say, that inconsistent assessments are inaccurate but after the approximation, they may reflect values that are useful for us. It must be stressed that there is no way to find the accurate values for the inconsistent input. 
By the definition, every approximation has an approximation error. The approximation error, in inaccurate data, is the discrepancy between an exact value and some approximation of it. The approximation error is commonly used in science and engineering as a percentage of what we approximate. By reducing inconsistencies, a clearly undesirable property in every system, the estimation error disappears since there is no approximation error for fully consistent PC matrices. The exact solution for the consistent PC matrix is the geometric mean which also happens to be the principal eigenvector. When normalized, they are identical.

\noindent The distance-based {\em inconsistency index} was proposed in \cite{Kocz93} as:

$$ii = \max (\min (|1-y/x/z|,|1-x\cdot z/y|))$$ 

\noindent for all triads $(x,y,z)$ specified by the consistency condition.

\noindent It has been simplified in \cite{KS2014a} to:

$$ii = \max (1-\min(\frac{y}{xz},\frac{xz}{y})),$$ 

\noindent which is equivalent to:

$$ii=\max(1- e^{-\left|\ln\left (\frac{y}{xz}\right )\right |}).$$

\noindent The expression $f(x,y,z)=|\ln(\frac{y}{xz})|$
is the distance of a triad $T$ from 0. When this distance increases, the $f(x,y,z)$ also increases. It is important to notice here that this definition allows us to localize the inconsistency in a PC matrix and it is of a considerable importance for most applications.  






For a given PC matrix $A$, let us denote the vector of arithmetic means of rows by $AM(A)$ and the vector of geometric means by $GM(A)$. 
A vector of geometric means of rows of a given matrix $A$ has a very natural interpretation.
It is transformed into a vector of arithmetic means by a logarithmic mapping.
The arithmetic mean has several properties that make it useful for a measure of central tendency.
Colloquially, measures of central tendency are often called averages.
For values $w_1,\ldots,w_n$, 
we have a mean $\bar{w}$, for which: $$(w_1-\bar{w}) + \ldots + (w_n-\bar{w}) = 0.$$
We may consider that the values to the left of the mean are balanced by the values to the right of the mean since 
$w_i-\bar{w}$ is the distance from a given number to the mean. The mean is the only value for which the residuals 
(deviations from the estimate) sum to zero.
When we are restricted to a single value for representing a set of known values $w_1,\dotsc,w_n$, 
the arithmetic mean may be used since it minimizes the sum of squared deviations 
from the typical value: the sum of $(w_i-\bar{w})^2$. 
In other words, the sample mean is also the best predictor
in the sense of having the lowest root mean squared error. Means were analyzed in \cite{Aczel1948} and in \cite{AczelSaaty1983}. 
It is difficult to state when exactly the logarithmic mapping was used for PC matrices as an alternative method for scaling priorities in hierarchical structures. However, it is usually attributed to Jensen. His results in \cite{Jensen84} occur in the same journal in which \cite{Saaty77} was published. Numerous subsequent Monte Carlo experiments have validated Jensen's findings (e.g., \cite{HK1996} used 1,000,000 matrices). 

For a matrix $A$ with positive coordinates, we define the matrix $B=\lambda(A)$ such that $b_{ij}=\ln(a_{ij})$. 
Reversely, for a matrix $B$, we define the matrix $A=\mu(B)$ such that $a_{ij}=\exp(b_{ij})$. 
By $\ln(x_1,\ldots,x_n)$ and $\exp(x_1,\ldots,x_n)$ denote vectors $(\ln(x_1),\ldots,\ln(x_n))$ 
and $(\exp(x_1),\ldots,\exp(x_n))$, respectively. It implies that:
\begin{equation}\label{ln}
\ln(GM(A))=AM(\lambda(A))
\end{equation}  and 
\begin{equation}\label{exp}
exp(AM(B))=GM(\mu(B)).
\end{equation}

\noindent If $A$ is consistent, then elements of $B=\lambda(A)$ satisfy 
$$b_{ij} + b_{jk}=b_{ik}$$ for every $i,j,k=1,2, \ldots ,n$. We call such a matrix {\em additively consistent}.

Let us consider an additive triad  $(x,y,z)$ (given or after the logarithmic transform). It is consistent if $y=x+z$ which is equivalent to the inner product of  $v=(x,y,z)$ by the vector $e=(1,-1,1)$, giving 0. This indicates that $v$ and $e$ are perpendicular vectors. In other words, if a triad is inconsistent, an orthogonal projection onto subspace perpendicular to the vector  $e=(1,-1,1)$ in the space $\mathbb{R}^3$ makes it consistent.
Such projections can be expressed by:
 $$ \tilde{v}=v - {v \circ e\over e\circ e} \, e,$$
 where $ u_1\circ u_2 $ is the inner 
 product of vectors $u_1$  and $u_2$ in $\mathbb{R}^3.$ \\

\begin{flushleft}
It holds: $$ e\circ e=3,\quad v \circ e= x-y+z,$$ hence $\tilde{v}=Av$, \\

\begin{flushleft}
where:
\end{flushleft}
\end{flushleft}
$$A=\left[
\begin{array}{rrr}
\frac{2}{3} & \frac{1}{3} & -\frac{1}{3} \\
\frac{1}{3} & \frac{2}{3} & \frac{1}{3} \\
-\frac{1}{3} & \frac{1}{3} & \frac{2}{3}
\end{array}
\right]
$$

\begin{flushleft}
{\em Remark}. A given inconsistent triad $(x,y,z)$ can be transformed into a consistent triad in the following three ways:
\end{flushleft}$$
(x,x+z,z),\quad
(y-z,y,z),\quad
(x,y, y-x).
$$

For a multiplicatively inconsistent triad $(a_{ik}, a_{ij}, a_{kj})$ its transformation to the consistent triad 
$(\tilde a_{ik}, \tilde a_{ij}, \tilde a_{kj})$ is given by:
$$
\tilde a_{ik}= a_{ik}^{2/3}a_{ij}^{1/3}a_{kj}^{-1/3},~~
\tilde a_{ij}= a_{ik}^{1/3}a_{ij}^{2/3}a_{kj}^{1/3},~~
\tilde a_{kj}= a_{ik}^{-1/3}a_{ij}^{1/3}a_{kj}^{2/3}.
$$

The above formulas were used in \cite{KKSX2015} for a Monte Carlo experimentation with the convergence of inconsistency by the sequence of inconsistency reductions of the most inconsistent triad.

\section{The geometric means and eigenvector}

The $n$-th step of the algorithm used in \cite{HK1996} transforms the most inconsistent triad $(a_{ij},a_{ik},a_{jk})=(x,y,z)$ of a given matrix $A_n$ into $\tilde{v}$ according to the formulas (\ref{ortho1}),(\ref{ortho2}),(\ref{ortho3}). Obviously, we also replace $(a_{ji},a_{ki},a_{kj})$ with $-\tilde{v}$, leaving the rest of the entries unchanged. Let $A_{n+1}$ denote the matrix after the transformation. The coordinates of $AM(A_n)$ may change only for the $i$-th, $j$-th and the $k$-th position. However, an elementary calculation shows that
\begin{eqnarray}
\tilde{x}+\tilde{y}&=& x+y\\
-\tilde{x}+\tilde{z}&=& -x+z\\
-\tilde{y}-\tilde{z}&=& -y-z
\end{eqnarray}.

This proves that 
\begin{equation} \label{AMn}
AM(A_n)=AM(A_{n+1}).
\end{equation}

\begin{figure}[h]
\centering
\includegraphics[width=0.9\linewidth]{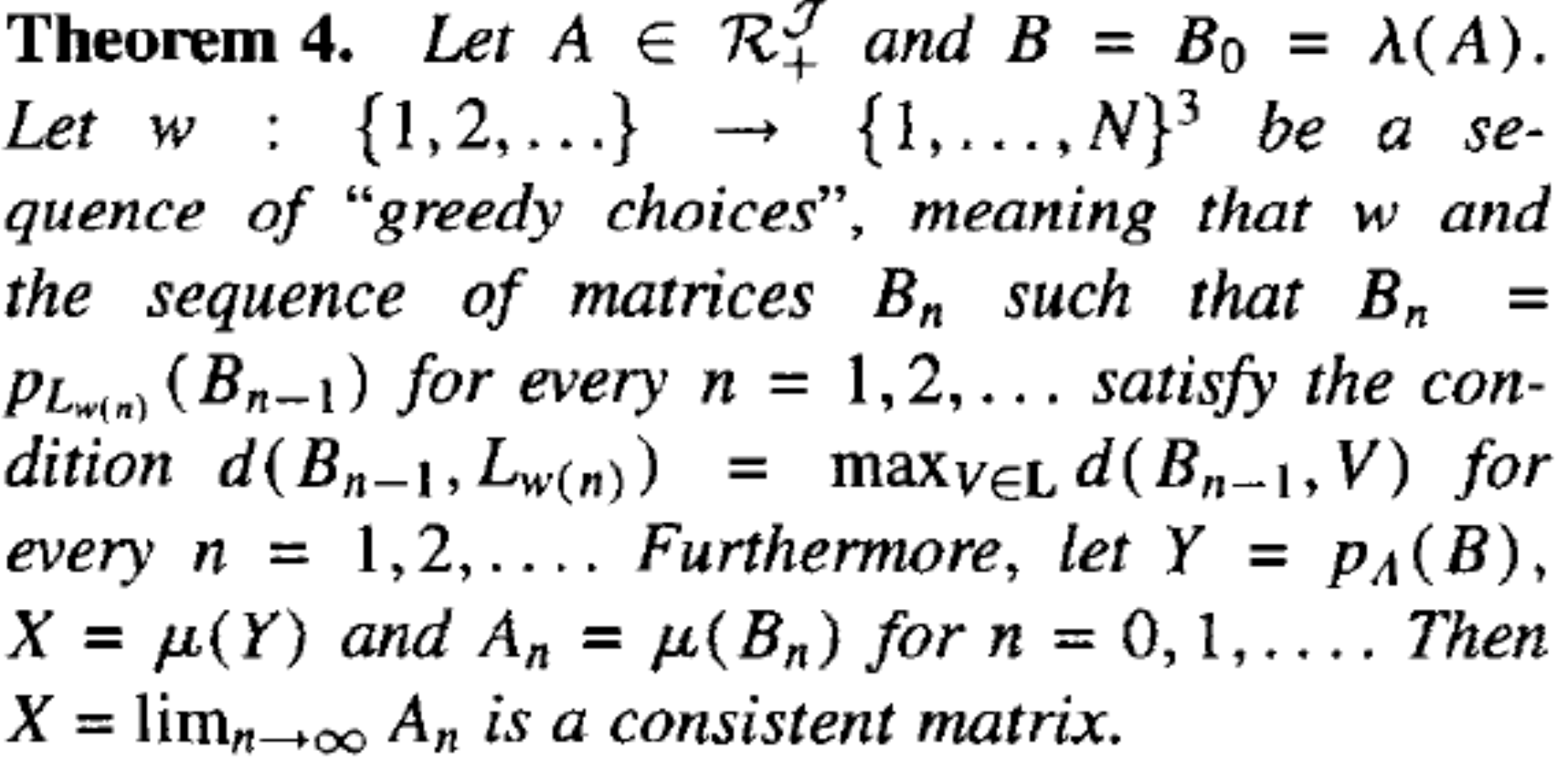}
\caption[Theorem 4]{Theorem 4 from \cite{HK1996}}
\label{fig:Th4}
\end{figure}

Theorem 4 in \cite{HK1996} (copied as it is for the traditional reasons and presented by Fig.~\ref{fig:Th4})
states that: $A_n$ is convergent to the orthogonal projection of $A_1$ onto the linear space of additively consistent matrices. From the above theorem and (\ref{AMn}), we get:

\begin{thm} \label{AM}
For a given matrix
\begin{displaymath}
A = \begin{bmatrix}
0 & a_{12} & \cdots & a_{1n} \\ 
-a_{12} & 0 & \cdots & a_{2n} \\ 
\vdots & \vdots & \vdots & \vdots \\ 
-a_{1n} & -a_{2n} & \cdots & 0
\end{bmatrix}
\end{displaymath}
and its orthogonal projection $A'$ onto the space of additively consistent matrices we have:
$$AM(A)=AM(A').$$
\end{thm}

The following two examples illustrate how the above works in practice.

\begin{ex}
Consider a classic PC matrix, introduced in \cite{Kocz93}:

$$
\begin{bmatrix}
1 & 2 & 5 \\ 
\frac{1}{2} & 1 & 3 \\ 
\frac{1}{5} & \frac{1}{3} & 1
\end{bmatrix} 
$$

After the orthogonal transform we get:

$$
\begin{bmatrix}
1 & 1.882072 & 5.313293 \\ 
0.531329 & 1 & 2.823108 \\ 
0.188207 & 0.35422 & 1
\end{bmatrix} 
$$


The first PC matrix is inconsistent since $2 \cdot 3 \ne 5$. The vector of geometric means in both cases is the same and equal to 

$$
\begin{bmatrix}
2.15443469 \\ 
1.14471424 \\ 
0.40548013
\end{bmatrix} 
$$

\end{ex}

\begin{ex}

Consider a matrix:

$$
\begin{bmatrix}
 0 & a & b & c \\ 
-a & 0 & d & e \\ 
-b & -d & 0 & f \\ 
-c & -e & -f & 0
\end{bmatrix} 
$$

The linear space of all such matrices is isomorphic with $\mathbb{R}^6$.

For such matrix to be additively consistent,
the following system of equation must hold:

$$U_1: a+d=b$$
$$U_2: d+f=e$$
$$U_3: a+e=c$$
$$U_4: b+f=c$$

Each equation describes a subspace $U_i$ of the dimension 5. Subspace $W$ of the consistent matrices is the intersection having dimension 3.

A matrix can be made consistent with the algorithm used in \cite{HK1996}. It corrects the most inconsistent triad, which corresponds to the orthogonal projection on $U_i$.
Another possible solution is the orthogonal projection on $W$. As a result, we obtain the matrix

$$ \begin{bmatrix}
 0 &  A &  B & C \\ 
-A &  0 &  D & E \\ 
-B & -D &  0 & F \\ 
-C & -E & -F & 0
\end{bmatrix}  ,
$$

where:
\begin{eqnarray*}
\begin{bmatrix}
A\\B\\C\\D\\E\\F
\end{bmatrix}
&=& \frac{1}{4}
\begin{bmatrix}
2a+b+c-d-e \\a+2b+c+d-f\\ a+b+2c+e+f\\ -a+b+2d+e-f\\ -a+c+d+2e+f\\ -b+c-d+e+2f
\end{bmatrix}.
\end{eqnarray*}
\end{ex}

Equations (\ref{ln}), (\ref{exp}) and Theorem \ref{AM} imply that for any reciprocal matrix $M$ to obtain the vector of weights of the closest consistent matrix, it is enough to calculate $GM(M)$.
In general, it is not so with the $EV(M)$, i.e. an eigenvector corresponding to the principal eigenvalue of $M$.

\begin{rem}
Assume
\begin{displaymath}
M = \begin{bmatrix}
1 & a & b \\ 
\frac{1}{a} & 1 & c \\ 
\frac{1}{b} & \frac{1}{c} & 1
\end{bmatrix}
\end{displaymath}
is a reciprocal matrix.
\begin{flushleft} Then \end{flushleft} $$GM(M)=EV(M)$$.
\end{rem}

\begin{proof}
Put \begin{displaymath}
v:=GM(M)= \begin{bmatrix}
\sqrt[3]{ab} \\ 
\sqrt[3]{\frac{c}{a}} \\ 
\frac{1}{\sqrt[3]{bc}}
\end{bmatrix}
\end{displaymath}

Then for $\lambda=1+\sqrt[3]{\frac{ac}{b}}+\sqrt[3]{\frac{b}{ac}}$, we have $Mv=\lambda v$ and it completes the proof.

\end{proof}

The above remark cannot be generalized to matrices of higher degrees.

\begin{ex}
Consider the matrix:

\begin{displaymath}
M = \begin{bmatrix}
1 & 2 & 1 & 3 \\ 
\frac{1}{2} & 1 & 1 & 1 \\ 
1 & 1 & 1 & 2 \\ 
\frac{1}{3} & 1 & \frac{1}{2} & 1
\end{bmatrix}
\end{displaymath}
Then
\begin{displaymath}
v:=GM(M)= \begin{bmatrix}
\sqrt[4]{6} \\ 
\sqrt[4]{\frac{1}{2}} \\ 
\sqrt[4]{2} \\ 
\sqrt[4]{\frac{1}{6}} \\ 
\end{bmatrix}
\end{displaymath}
and there is no $\lambda$ such that $Mv=\lambda v$, and it follows that $GM(M)\neq EV(M)$. Vector $GM(M)$ generates the consistent matrix which is the closest to $M$. Consequently, $EV(M)$ does not.
\end{ex}

\begin{figure}[h]
\centering
\includegraphics[width=0.75\linewidth]{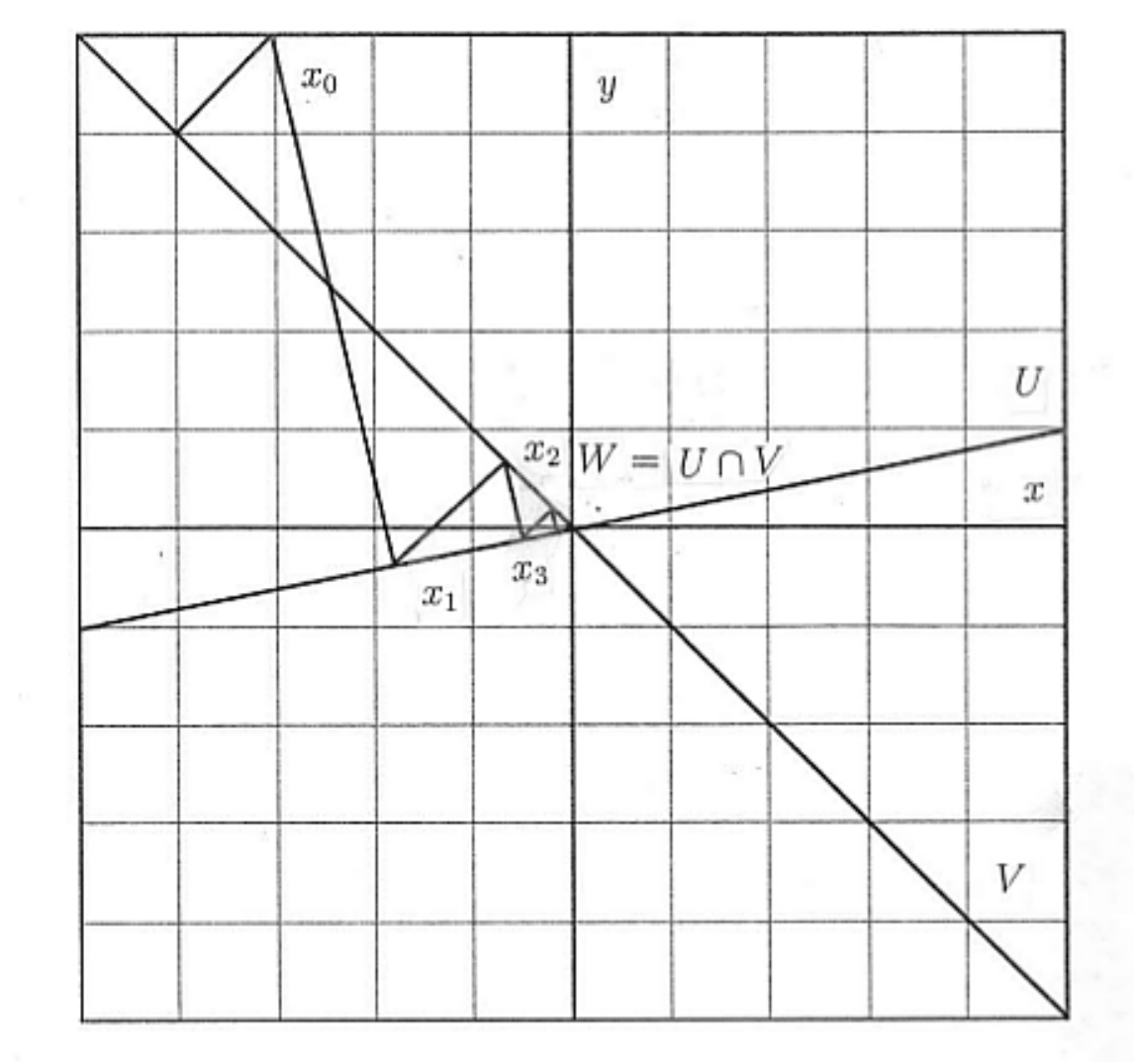}
\caption[Orthogonal projections]{Orthogonal projections}
\label{fig:OrthogProj}
\end{figure}

For a given inconsistent PC matrix $A$, we can find a consistent PC matrix by one transformation which is evidently  $[v_i/v_j]$ and $v_i=GM_i(A)$ ($GM_i(A)$ = geometric mean of the i-th row of $A$). The triad-by-triad reduction for $n=3$ is simple since there is only one triad hence only one transformation is needed. Orthogonal transformations may be indefinite for $n>3$ but the convergence is very quick, as demonstrated by \cite{KKSX2015} and usually, fewer than 10 steps are needed in most applications.

\begin{figure}[h]
\centering
\includegraphics[width=0.95\linewidth]{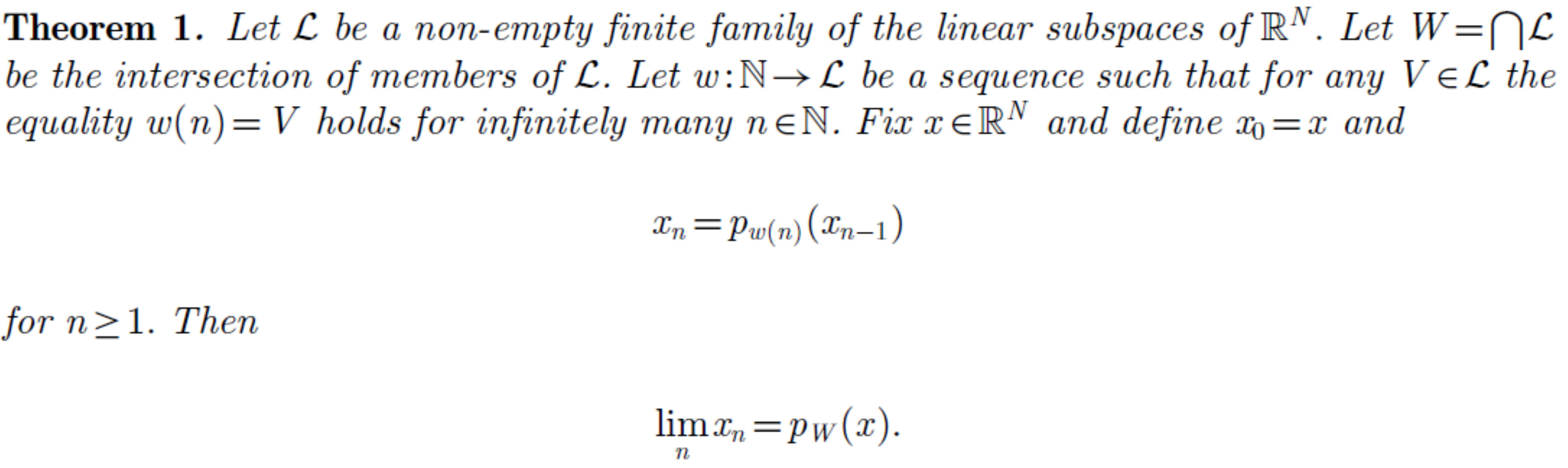}
\caption[Theorem 1]{Theorem 1 from \cite{KS2010}}
\label{fig:Th1}
\end{figure}

For an inconsistent matrix, we find the most inconsistent triad (according to $ii$) and by using the three expressions for the orthogonal projection, we make it consistent. The idea of the algorithm is shown in Fig. \ref{fig:OrthogProj}. 
We only need one step for a PC matrix $3 \times 3$ since there is only one triad and changing it does not influence ``other triads''. When $n>3$, changes  to all three values in one triad propagate to other triads. It is not evident for $n\rightarrow \infty$ how the propagation may go. However, the proof of convergence was provided independently in \cite{BB1996} and in \cite{KS2010} and resulted in Theorem 1 (Fig. \ref{fig:Th1}).

Eventually, we get a consistent matrix which can be obtained from a vector of geometric means of rows of the original inconsistent matrix.
The result is of considerable importance since the principal eigenvector of the input matrix is assumed to be the only  solution according to \cite{Saaty77} but it is in the direct contradiction with the mathematical proofs provided here.

The presented results in this study are also consistent with findings in \cite{BV2004}, explicitly stating in its conclusions that:

\begin{quotation}
Our main conclusion
is that, although the EM is very elegant from
a mathematical viewpoint, the priority vector
derived from it can violate a condition of order preservation
that, in our opinion, is fundamental in
decision aiding -- an activity in which it is essential
to respect values and judgments. In light of that,
and independently of all other criticisms presented
in the literature, we consider that the EM has a serious
fundamental weakness that makes the use of
AHP as a decision support tool very problematic.
\end{quotation} $EM$ in the quoted text stands for ``eigenvector method''.

\section{Conclusions}

This study finally closes the ongoing (since 1983 when \cite{Jensen84} was published) dispute whether the principal eigenvector or a vector of geometric means is a better approximation for an inconsistent PC matrix.
For fully consistent pairwise comparisons matrices, such dispute has no scientific merits since the principal eigenvector $EV$ is equal to the vector $GM$ of geometric means of rows. For inconsistent pairwise comparisons, numerous Monte Carlo experiments (starting with the first study \cite{Jensen84} and with 1,000,000 analyzed cases in \cite{HK1996b}) demonstrated that, for the Euclidean distance, $EV-$generated approximation of $A$ is (on average) not as close as the one generated by $GM$.

The difference between $GM$ and $EV$ approximations is not of noticeable significance from the application point of view for ``not-so-inconsistent'' (NSI) PC matrices (NSI concept was introduced in \cite{HK1996}). If the difference is small, why bother? The difference between Newton's law of universal gravitation and the theory of relativity is small yet one can miss a planet. 
In fact, the accuracy of approximation is not really of great importance considering that the input (ofter highly subjective) may be inconsistent and plus/minus 10\% accuracy would be more of a dream than reality. Often, $EV$ and $GM$  differ by the third or second decimal digit while the significance of even the first decimal digit of the input data may be questioned. 
It is the inconsistency improvement which really contributes to the approximation accuracy improvement. For this reason, the convergence of the inconsistency improvement to the vector of geometric means; not to the principal eigenvector, is of the fundamental importance, not only to finally conclude the ongoing (since 1984) dispute what is (or not) better but also for practical applications. Geometric means can be computer with a pocket calculator (Gnumerics and Excel are even easier to use) while computing the principal eigenvector is not as easy as computing geometric means. An eigenvalue perturbation problem exists and finding the eigenvectors and eigenvalues of a system that is perturbed from known eigenvectors and eigenvalues is not a trivial problem to handle.

It is worth noticing that the simplified version of pairwise comparisons, presented in \cite{KS2015a} does not have inconsistencies. PC matrix elements are generated from a set of principal generators, preserving the consistency condition.

The pairwise comparisons method has been implemented, as a part of the cloud computing support, for a group decision making process used by software development team at Health Sciences North (a regional hospital in Sudbury, Ontario). The software is available for downloading from SourceForge.net). With more than 400K projects and over 3 million registered users, it is one of the biggest repositories in the world of open source software development projects.

\section*{Acknowledgment}
The research of the first author has been partially supported by the Provincial Government through the Northern Ontario Heritage Fund Corporation and by the Euro Grant Human Capital. The research of the second author has been partially supported by the Polish Ministry of Science and Higher Education. 

The authors would like to thank Professor Ryszard Szwarc for his contribution in proving mathematical proofs to formulas for the orthogonal projection. Authors would like to thank Amanda Dion-Groleau (Laurentian University student) and Grant O. Duncan (a part-time graduate student at Laurentian University and Team Lead, Business Intelligence and Software Integration, Health Sciences North, Sudbury, Ontario) for their help with proofreading this text.


\begin{thebibliography}{9}


\bibitem{Aczel1948}
Aczel, J.,
On means values. Bulletin American Mathematical Society 54: 392-400, 1948.

\bibitem{AczelSaaty1983}
Aczel, J., Saaty, L.T.,
Procedures for Synthesizing Ratio Judgements,
Journal of Mathematical Psychology 27: 93--102, 1983.

\bibitem{BV2004}
Bana e Costa, C.A., Vansnick, J-C., 
A critical analysis of the eigenvalue method used to derive priorities in AHP,
European Journal of Operational Research, 187(3): 1422-142, 2004.


\bibitem{BB1996}
Bauschke, H.H, Borwein, J.M.,  On Projection Algorithms for Solving Convex Feasibility Problems, SIAM Rev. 38(3): 367-426, 1996.


\bibitem{GS1958}
Gerard, HB, Shapiro, HN, Determining the Degree of Inconsistency in a Set of Paired Comparisons, Psychometrika,  23(1): 33-46, 1958

\bibitem{HK1996b}
Herman, M.W., Koczkodaj, W.W.,
A Monte Carlo study of pairwise comparison,
Information Processing Letters,57(1): 25-29, 1996.

\bibitem{H1953}
Hill, RJ, A Note on Inconsistency in Paired Comparison Judgments,
American Sociological Review, 18(5): 564--566, 1953.

\bibitem{HK1996}
Holsztynski, W., Koczkodaj, W.W., \emph{Convergence of Inconsistency
Algorithms for the Pairwise Comparisons}, IPL. 59(4): 197-202, 1996.

\bibitem{Jensen84}
Jensen, R. An Alternative Scaling Method for Priorities in Hierarchical Structures,
Journal of Mathematical Psychology, 28: 317-332, 1984.

\bibitem{KSB1939}
Kendall, M.G., Smith, B., On the Method of Paired Comparisons,
Biometrika, 31(3/4): 324-345, 1940.

\bibitem{Kocz93}
Koczkodaj, W.W., A New Definition of Consistency of Pairwise Comparisons, 
Mathematical and Computer Modelling, 18(7): 79-84, 1993.

\bibitem{KKSX2015}
Koczkodaj, W.W., Kosiek, M., Szybowski, J. and Xu, D., Fast Convergence of Distance-based Inconsistency in Pairwise Comparisons, Fundamenta Informaticae, (in print)




\bibitem{KS2010} Koczkodaj, W.W., Szarek, S.J., On distance-based inconsistency reduction algorithms for pairwise comparisons, Logic J. of the IGPL, 18(6): 859-869, 2010.

\bibitem{KS2014a}
Koczkodaj, W.W., Szwarc, R., On Axiomatization of Inconsistency Indicators for Pairwise Comparisons, Fundamenta Informaticae 132(4): 485-500,  (online: arXiv:1307.6272v4), 2014.

\bibitem{KS2015a}
Koczkodaj, W.W., Szybowski, J., Pairwise comparisons simplified, Appl. Math. Comput. 253: 387--394, 2015.

%



\bibitem{Saaty77}  
Saaty, T.L. {\it A Scaling Method for Priorities in Hierarchical Structure}, Journal of Mathematical Psychology 15: 234-281, 1977.

\bibitem{S1961}
Slater, P., Inconsistencies in a Schedule of Paired Comparisons
Biometrika 48(3/4): 303-312,  1961.

\bibitem{XDXW2008}
Xu, WJ , Dong, YC, Xiao, WL, Is It Reasonable for Saaty's Consistency
Test in the Pairwise Comparison Method? in Proceedings of 2008 ISECS
International Colloquium on Computing, Communication, Control, and
Management 3: 294-298, 2008.



\end{thebibliography}
\end{document}